\documentclass[10pt,letterpaper]{article}
\usepackage{opex3}

\begin{document}

\title{Enhanced transmission of transverse electric waves through periodic arrays of structured subwavelength apertures}

\author{Sanshui Xiao, Liang Peng, and Niels Asger Mortensen}

\address{DTU Fotonik - Department of Photonics Engineering, \\Technical
University of Denmark, DK-2800 Kongens Lyngby, Denmark.}

\email{saxi@fotonik.dtu.dk} 



\begin{abstract}
Transmission through sub-wavelength apertures in perfect metals is expected to be strongly suppressed. However, by structural engineering of the apertures, we numerically demonstrate that the transmission of transverse electric waves through periodic arrays of subwavelength apertures in a thin metallic film can be significantly enhanced.
Based on equivalent circuit theory analysis, periodic arrays of square structured subwavelength apertures are obtained with a 1900-fold transmission enhancement factor when the side length $a$ of the apertures is $10$ times smaller than the wavelength ($a/\lambda=0.1$). By examining the induced surface currents and investigating the influence of the lattice constant and the incident angle to the resonant frequency, we show that the enhancement is due to the excitation of the strong localized resonant modes of the structured apertures.
\end{abstract}



\section{Introduction}

Transmission through a sub-wavelength aperture in a perfect metal is expected to be strongly suppressed. This follows from Bethe's theoretical description and Rayleigh's criterion~\cite{Bethe1944,Roberts1987}, which shows that, for a single subwavelength aperture (with
a radius of $r$) perforated in a thin perfectly conducting film, the transmission is proportional to $(r/\lambda)^4$
and is quite weak when $r\ll \lambda$. Here, $\lambda$ is the free-space wavelength of the incident radiation.
Since the first experimental demonstration of the extraordinary transmission through periodic arrays of subwavelength holes milled in optically thick metallic films~\cite{Ebbesen1998P1}, there has been a renewed and considerable interest for the phenomenon. Although the physical mechanism behind the extraordinary transmission has been heavily debated, numerous theoretical and numerical studies support that the enhanced transmission is attributed to the surface plasmon resonance arising from
the surface periodicity and the localized resonances of the subwavelength apertures~\cite{Ghaemi1998P1,Porto1999P1,Martin2001P1,Krishnan2001P1,Ruan2006,Xiao2007P2}.
Extraordinary transmission has been extensively studied in periodic subwavelength hole arrays, including systems with corrugated metal surfaces~\cite{Bai2005P1,Giannattasio2004P1,Bonod2003P1,Avrutsky2000P1,Xiao2007P1} and different aperture geometries~\cite{Ruan2006,Garcia2005P1}. Most recently, the extraordinary transmission phenomena has also been explored in systems where the aperture is in close proximity to a metamaterial structure~\cite{Aydin2009P1,Cakmak2009P1,Ye2009P1}. It has been shown that the transmission through subwavelength apertures is in general strongly polarization-dependent. For regular rectangular subwavelength apertures~\cite{Ruan2006} with the case of normal incidence, enhanced transmission is observed only when the electric field is polarized along the short axis of the apertures, so-called TM polarization. However, the transmission is quite weak for TE polarization, i.e., the electric field of the incident wave is polarized along the long side of the apertures.
While there is a consensus that surface-plasmon resonances are in one or the other way responsible for the enhanced transmission at optical frequencies (where metals are dispersive and display plasmonic behavior), it is not fully clear how to extend the phenomenon to longer wavelengths where the metal eventually approaches the perfectly conducting regime. While surface plasmons are not easily supported in e.g. the microwave regime, artificial structuring of a metal surface allows the formation of spoof surface plasmon even on the surface of ideal metals~\cite{Pendry2004P1,Hibbins2005P1}.
In this paper, we propose that enhanced transmission of TE polarized radiation may be achieved by structuring the subwavelength aperture in an ideal metal film. With the aid of the equivalent circuit theory, we demonstrate extremely high enhancement factor (1900-fold) when the side length of each structured aperture is $10$ times smaller than the free space wavelength.
\begin{figure}[htbp]
\centering\includegraphics[width=13cm]{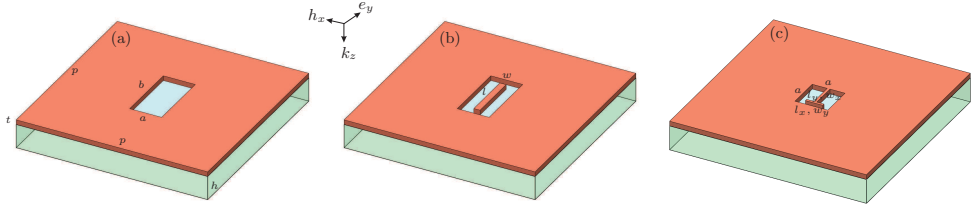}
\caption{Unit cell for a periodic array of subwavelength (a) regular rectangular apertures; (b) rectangular apertures incorporated with metallic bars; (c) square apertures incorporated
with T-shape metallic bars. Axis shows the propagation direction and polarization of the incident plane wave.}
\label{structure}
\end{figure}

\section{Enhanced transmission with structured subwavelength aperture arrays}
\begin{figure}[htbp]
\centering\includegraphics[width=13 cm]{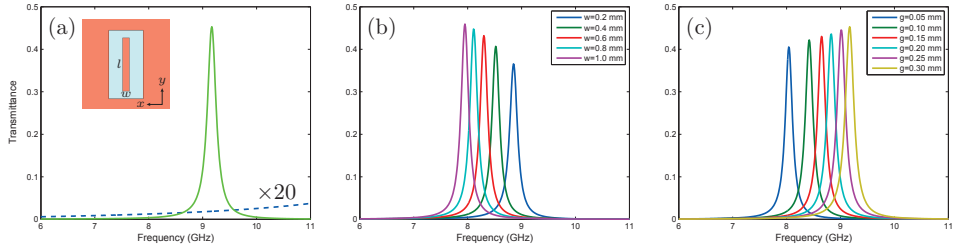}
\caption{(a) Transmittance spectra for periodic arrays of subwavelength apertures with (the green solid line) and without (the blue dashed line) the metallic bars. The inset shows the top-view of the central part of the structure with the metallic bars as shown in Fig.~1(b). Note that the lower trace has been magnified by a factor of 20. (b) Transmittance as a function of the width of the metallic bars $w$. (c) Transmittance versus the distance $g$ between the bars and the short axis of the subwavelength apertures, where $g=(b-l)/2$ ($b$ and $l$ is the length of the apertures and of the metallic bars in the $y$ direction, respectively).}
\label{rectangular}
\end{figure}

We imagine that a two dimensional square periodic array of subwavelength holes is perforated along the $z$ direction
in a copper film deposited on a FR4 dielectric substrate.
The unit cell of the corresponding structure is illustrated in Fig.~\ref{structure} (a), where in this
paper the thickness $t$ of the copper layer and the thickness $h$ of the FR4 substrate are fixed as
$t$=30 $\mu$m and $h$=0.5~mm, respectively. The period of the array is $p$=30~mm, and the side length of
the rectangular subwavelength hole is denoted by $a$ and $b$ in the $x$ and $y$ direction, respectively.
Here, we consider the normally incident case of the TE plane wave, i.e., the electric
field of the incident wave is polarized along the $y$ direction, as indicated in Fig.~\ref{structure}. To illustrate the transmission properties we utilize fully vectorial simulations of the Maxwell equations, performed with the aid of CST MICROWAVE STUDIO~\cite{CST}.
The electric constant and tangent loss of the FR4 dielectric substrate are taken as $\epsilon=3.6$ and
$\delta=0.01$, respectively. We first study the transmission through a periodic array of the regular rectangular subwavelength holes
with $a$=5~mm and $b$=10~mm, and the result is shown by the dashed line in
Fig.~\ref{rectangular}(a). Notice that the values have been magnified by a factor of 20 for clarity.
We emphasize that the transmittance is quite low, less than 0.2$\%$, even for such a thin film at microwave frequencies.
Note that the cutoff frequency for such a rectangular waveguide is 15~GHz.
When the frequency is below the cutoff frequency, only evanescent modes can be excited. Obviously the transmission relying on the evanescent tunneling process is quite weak, as shown by the dashed line in Fig.~\ref{structure}(a).
Actually, we observe high transmittance (not shown here) for the TM polarized plane wave in the frequency of our interest,
where the frequency is below the cutoff.
We believe that in this case the enhanced transmission is associated to the excitation of the so-called spoof surface plasmons~\cite{Ruan2006,Pendry2004P1,Hibbins2005P1,Shen2008P1}. Most recently, it has been shown that the electric field of the spoof surface plasmon is mainly along the short axis of the rectangular apertures~\cite{Ruan2007P1}. Therefore, this kind of spoof surface plasmon resonance can only be excited by the TM polarized plane wave, thus leading to the enhanced transmission. In this paper, we mainly focus on how to realize high transmission through arrays of subwavelength apertures for the TE polarized plane wave when the working frequency is below the cut-off of the waveguide mode.

In order to increase the transmission for the TE plane wave, we use artificial structures to create a resonance state below the cut-off of the waveguide mode. In our proposal, a periodic array of copper bars has been added to the centers of the subwavelength holes, as shown in Fig.~\ref{structure}(b). To see this more clearly, a top-view of the central part of the corresponding structure is illustrated in the inset of Fig.~\ref{rectangular}(a). The position and the thickness of the metallic bars in the $z$ direction are exactly same as the film, thus such a composite structure can be realized experimentally without adding more steps in the etching process. The width in the $x$ direction and the length in the $y$ direction for the metallic bars are denoted by $w$ and $l$, respectively. The solid green line in Fig.~\ref{rectangular}(a) represents the transmittance spectrum when $w$=0.5~mm and $g=(b-l)/2$=0.3~mm, where $g$ is the distance between the boundaries of the bars and the short axis of the subwavelength apertures in the $y$ direction. Other parameters are exactly the same as those used for the result shown by the dashed line in Fig.~\ref{rectangular}(a). In Fig.~\ref{rectangular}(a), one can observe a strong resonance transmission peak at the frequency of $f$=9.17~GHz ($\lambda$=32.71~mm) with the transmittance up to 40$\%$, which is about 600 times larger than that for the case without the metallic bars. Our interpretation is that the enhanced transmission is attributed to a resonant effect resulting in the effective coupling of light into subwavelength apertures. Influence of the geometric parameters of the metallic bars to the transmittance spectrum is investigated in the following. Figure~\ref{rectangular}(b) shows the transmittance spectrum as a function of the width of the metallic bars when $g$=0.1~mm. The maximum value of the transmittance does not vary significantly when increasing $w$. On the other hand, the resonance red-shifts. When $w$ is fixed at 0.5~mm, the transmittance almost maintains the same level while shortening the gap $g$ causes a red-shift, as shown in Fig.~\ref{rectangular}(c).

\begin{figure}[htbp]
\centering\includegraphics[width=7 cm]{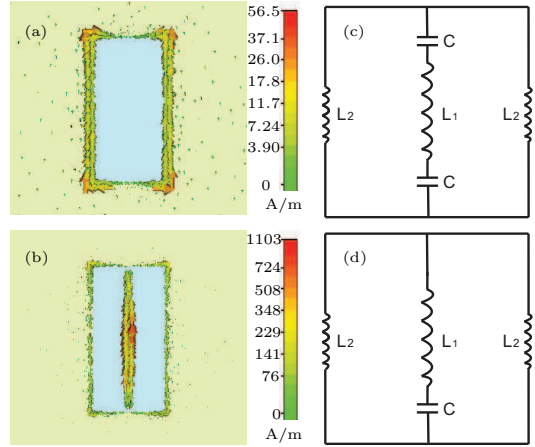}
\caption{The induced surface currents  for the subwavelength apertures without (a) or with (b) the metallic bars at the central plane of the film structure in the $z$ direction. (c) Equivalent LC circuit for the result shown in (b). (d) Modified equivalent LC circuit.}
\label{current}
\end{figure}

To better understand the mechanism behind the enhanced transmission, we next explore the induced surface currents. Figures~\ref{current}(a) and~\ref{current}(b) show results for the subwavelength apertures without and with the metallic bars at the central plane of the film structure in the $z$ direction, respectively. The frequency corresponds to the resonant frequency $f$=9.17~GHz for the structure shown in Fig.~\ref{structure}(b). Notice that strong induced currents appear especially in the vicinity of the subwavelength apertures.
One can see from Fig.~\ref{current}(a) that  the surface currents mostly remain at the surfaces of the subwavelength aperture, with high current densities at the discontinuous corners of the subwavelength aperture. Compared with the results shown in Fig.~\ref{current}(b), the induced surface currents shown in Fig.~\ref{current}(a) are quite weak, thus leading to a weak contribution to the radiation.
It turns out that the surface currents are significantly amplified for the subwavelength aperture with metallic bar structures, indicated by the color bars in
Fig.~\ref{current}. The composed structure strongly concentrates the induced currents, which is associated with a resonance behavior.
Here we present an equivalent circuit, as shown in Fig.~\ref{current}(c) for the structure shown in Fig.~\ref{structure}(b). The circuit theory predicts that the resonant frequency is equivalent to
\begin{eqnarray}
\omega=\frac{1}{\sqrt{(L_1+L_2/2)C/2}},
\end{eqnarray}
where $L_1/L_2$ and $C$ represent the capacitance and inductance, respectively. From Fig.~\ref{current}(b), one can assume $L_2$ is much smaller than $L_1$. The values of the capacitance at the two gaps are identical. As mentioned above, the main goal of this paper is to realize subwavelength  aperture structures with high transmission and working in a low frequency. To lower the resonant frequency, obviously one can increase the values of $L_1$ and $C$. For the structure studied, one knows that the value of $C$ is proportional to $w$ (the width of the bars in the $x$ direction) and inversely proportional to $g$ (the size of the gap in the $y$ direction), and that the value of $L$ is proportional to $l$ (the length of the bars in the $y$ direction) and is almost independent on $g$ since $g\ll l$.
From the equivalent circuit theory, it is known that the resonant frequency drops down with decreasing of $g$ when $w$ is fixed, which is in agreement with the result in Fig.~\ref{rectangular}(c). Likewise, $C$ increases while $L_1$ drops down with increasing of $w$. In this way we conclude that the increasing of $C$ is dominant for lowering the resonant frequency since we observe the red-shift in frequency from Fig.~\ref{rectangular}(b).
We furthermore note that one may easily decrease the resonant frequency by closing one of the gaps in the $y$ direction, as shown in Fig.~\ref{current}(d). In this situation, the resonant frequency changes to $\omega=1/(\sqrt{(L_1+L_2/2)C})$, which is a factor of $\sqrt{2}$ times smaller than before.

\begin{figure}[htbp]
\centering\includegraphics[width=13 cm]{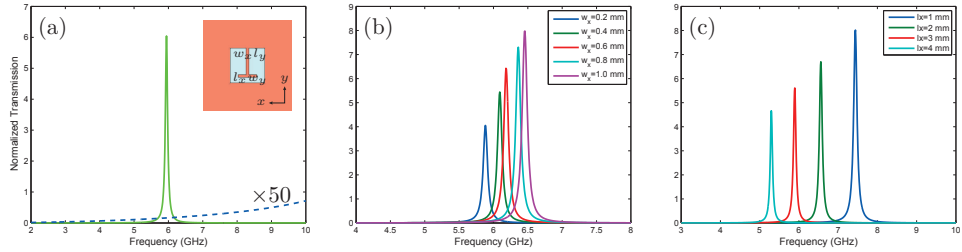}
\caption{(a) Normalized transmission for aperture structures with (the green solid line) and without (the blue dashed line) the T-shape metallic bars. The inset shows the top-view of the central part of the structure with the T-shape bars as shown in Fig. 1(c). Note that the lower trace has been magnified by a factor of 50. (b) Normalized transmission as a function of $w_x$ when $l_x$=2.5~mm, $g$=0.1~mm and $w_y$=0.03~mm. (c) Normalized transmission versus $l_x$ when $w_x$=0.5 ~mm, $g$=0.1~mm and $w_y$=0.03~mm.
The bar in the $y$ direction are described by $w_x$ and $l_y$ and the bar in the $x$ direction are denoted by $l_x$ and $w_y$. $g$ is defined as $g=(a-l_y-w_y)$, where $a$ is the side length of the square aperture. }
\label{square}
\end{figure}

In order to miniaturize the subwavelength aperture and further lower the resonant frequency, we next investigate the structure by closing one of the gaps as shown in Fig.~\ref{structure}(c), where the square subwavelength hole has a side length $a$ of 5~mm.
Notice that in this case the cutoff frequency for such a square hollow waveguide is 30~GHz.
The parameters for the thickness of copper layers, the lattice constant and the thickness of the substrate are exactly the same as those used above. Now we characterize the transmittance utilizing the widely-used concept "normalized transmission", which is defined as
\begin{eqnarray}
T_{\rm norm}=\frac{T}{S_{\rm aperture}/S_{\rm cell}}=\frac{P_{\rm out}}{P_{\rm in}}\frac{S_{\rm cell}}{S_{\rm aperture}},
\end{eqnarray}
where $P_{\rm in}/P_{\rm out}$ is the power flux through in/out the metal film, $S_{\rm aperture}$ is the aperture area in a unit cell, and
$S_{\rm cell}$ is the area the unit cell. $T_{\rm norm}$ should be larger than unity for the enhanced transmission.
Figure~\ref{square}(a) shows the normalized transmission for square aperture structures without (the blue dashed line) the T-shape bars and with (the green solid line) the bars when $w_x$=0.5~mm, $g$=0.1~mm, $l_x$=3~mm, and $w_y$=0.03~mm. The bar in the $y$ direction has dimensions $w_x$ and $l_y$ and likewise the bar in the $x$ direction has dimensions denoted by $l_x$ and $w_y$, respectively. Note that $g$ is defined as $g=(a-l_y-w_y)$. In the results, a strong resonance transmission peak appears at the frequency of 5.94~GHz ($\lambda$ =5.05~cm) with the normalized transmission of around 6. Compared with the case without the T-shape bars, the normalized transmission is extremely enhanced, in this case by a factor of 1900. For this particular example, the side length $a$ of the square apertures is around $10$ times smaller than the resonant wavelength.
Figures~\ref{square}(b) and~\ref{square}(c) show how the normalized transmission and the resonant frequency change as we tune the T-shape structure.
With decreasing of $w_x$ or increasing of $l_x$, the resonant frequencies red-shift, which is in consistence with our expectations based on the equivalent circuit theory. However, the normalized transmission peaks rapidly drop when we decrease $w_x$ or increase $l_x$. Obviously, it remains a challenging and open question how to optimally design a complex structure, where the resonant frequency is relatively low and the normalized transmission is quite strong.

\begin{figure}[htbp]
\centering\includegraphics[width=10 cm]{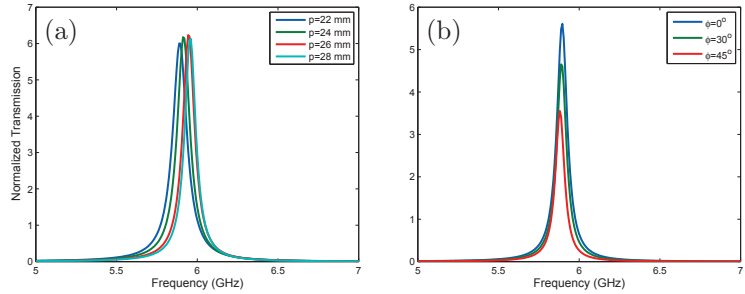}
\caption{Normalized transmission as a function of (a) the lattice constant $p$ (b) the incident elevation angle $\phi$.}
\label{tuning}
\end{figure}

We emphasize that there is a continuing discussion about the physical mechanisms responsible for the extraordinary
transmission~\cite{Ghaemi1998P1,Porto1999P1,Martin2001P1,Krishnan2001P1,Ruan2006}.
It has been shown that the enhanced transmission
is attributed to the surface plasmon resonance arising from
the surface periodicity~\cite{Ghaemi1998P1,Porto1999P1,Martin2001P1,Krishnan2001P1}.
The resonance appears when the wavevector of the surface plasmon matches the wave vector of the incident photon and grating as follows:
$k_{\rm sp}=k_x\pm n G_x\pm mG_y$,
where $k_{\rm sp}$ is the surface plasmon wavevector, $k_x=(2\pi/\lambda) sin(\theta)$ is the component of the incident wavevector in the plane of the grating, and
$G_x=G_y=2\pi/p$ is the grating reciprocal vector for the square array. Apparently this kind of resonance is strongly dependent on the period and the incident angle, while it is almost independent on the hole shape.
More recently, it was found that there is a strong influence of the hole shape on the enhanced transmission, arising from the localized resonances~\cite{Klein2004P1,Van2005}. Obviously, the localized resonances are mainly determined by the aperture's shape, independent on the period, as well as the incident angle. As mentioned above, the spoof surface plasmon resonance arisen from the periodicity can not be excited by the TE polarized plane wave. Beside, from Fig.~\ref{current}(b) we almost know that the enhanced transmission is mainly due to the excitation of the localized resonance of the composed apertures. In order to further support it, we here investigate how the resonant frequency changes when tuning the lattice constant $p$ of the structure and the incident elevation angle $\phi$, as plotted in Fig.~\ref{tuning}.
It shows that the transmission peak is almost independent on the lattice constant and the incident elevation angle, which
indicates that the resonance is the localized resonance and that Bloch surface waves are not contributed to the enhanced transmission. In such a structure, surface waves are not expected to play a significant role at frequencies where the metal is close to being an ideal metal. Due to weak coupling of the neighboring resonators, it is naturally understood that a small shift of the resonant frequency is observed in Fig.~\ref{tuning}(a) when tuning the lattice constant. In Fig.~\ref{tuning}(b) one sees that the normalized transmission drops down when increasing the elevation angle.
The overlap between the localized resonant mode and the incident wave becomes weaker when increasing the elevation angle, thus resulting in low transmission.

\section{Conclusion}
The transmission of the TE polarized plane wave through a periodic array of the regular subwavelength apertures in a thin metallic film is quite weak when the frequency is below the cut-off of the waveguide mode. In this paper, we have systemically investigated how to increase the transmission by structural engineering of the aperture structures. The transmission can be significantly enhanced with a 1900-fold factor when we have added T-shape metallic bars into the aperture structures with the side length $10$ times smaller than the resonant wavelength. The resonant frequencies vary when tuning the composite structures,
which can be well explained by the equivalent circuit theory. We have shown that the physical mechanism for enhancement is attributed to the excitation of the strong localized resonance of the structured apertures by examining the induced surface currents and investigating the influence of the lattice constant and incident angle to the resonant frequency.

\section*{Acknowledgments}
This work is financially supported by the Danish Research Council for Technology and Production Sciences (grant no: 274-07-0379) as well as the Danish National Advanced Technology Foundation (grant no: 004-2007-1).

\end{document}